\documentclass[showpacs,aps,prl,twocolumn,superscriptaddress]{revtex4}

\usepackage{graphicx}
\usepackage{amsmath}

\begin{document}

\title{Electronic transport and vibrational modes in the smallest molecular 
bridge: H$_2$ in Pt nanocontacts}

\author{Y.~Garc\'{\i}a}
\affiliation{Departamento de F\'{\i}sica Aplicada, Universidad de
Alicante, San Vicente del Raspeig, Alicante 03690, Spain.}
\author{J.~J.~Palacios}
\affiliation{Departamento de F\'{\i}sica Aplicada, Universidad de
Alicante, San Vicente del Raspeig, Alicante 03690, Spain.}
\affiliation{Unidad Asociada UA-CSIC, Universidad de
Alicante, San Vicente del Raspeig, Alicante 03690, Spain.}
\author{E. SanFabi\'an}
\affiliation{Departamento de Qu\'{\i}mica-F\'{\i}sica, Universidad de
Alicante, San Vicente del Raspeig, Alicante 03690, Spain.}
\affiliation{Unidad Asociada UA-CSIC, Universidad de
Alicante, San Vicente del Raspeig, Alicante 03690, Spain.}
\author{J. A. Verg\'es.}
\affiliation{Departamento de Teor\'{\i}a de la Materia Condensada, Instituto
de Ciencias de Materiales de Madrid ICMM-CSIC, Cantoblanco, 
Madrid 28049, Spain.}
\author{A. J. P\'erez-Jim\'enez}
\affiliation{Departamento de Qu\'{\i}mica-F\'{\i}sica, Universidad de
Alicante, San Vicente del Raspeig, Alicante 03690, Spain.}
\author{E. Louis}
\affiliation{Departamento de F\'{\i}sica Aplicada, Universidad de
Alicante, San Vicente del Raspeig, Alicante 03690, Spain.}
\affiliation{Unidad Asociada UA-CSIC, Universidad de
Alicante, San Vicente del Raspeig, Alicante 03690, Spain.}

\date{\today}

\begin{abstract}
We present a state-of-the-art first-principles analysis of 
electronic transport in a Pt nanocontact in the presence of H$_2$
which has been 
recently reported by Smit {\em et al.} in Nature {\bf 419}, 906 (2002). 
Our results indicate that at the last stages of the breaking of the 
Pt nanocontact two basic forms of bridge involving H can appear.
Our claim is, in contrast to Smit {\em et al.}'s,
that the main conductance histogram peak at $G\approx 2e^2/h$ is 
not due to molecular H$_2$, 
but to a complex Pt$_2$H$_2$ where the H$_2$ molecule
dissociates. A first-principles
vibrational analysis that compares favorably with the experimental one
also supports our claim. 
\end{abstract}

\pacs{73.63.Fg, 71.15.Mb}

\maketitle

Prediction ability is, possibly, the most important 
attribute associated with first-principles numerical implementations
of quantum transport theory in molecular and atomic-scale systems. 
This desirable property relies on the correct handling of two closely-related 
factors: The atomic structure and the electronic structure of the
nanoconstriction or molecular junction. The calculation
of the electronic
structure in current-carrying situations is a topic of increasing interest and 
has received due
attention in recent years\cite{Lang:prb:95,Yaliraki:jcp:99,
Taylor:prb:01:b,Palacios:prb:01,Palacios:prb:02,Damle:prb:01,Brandbyge:prb:02}.
The computation of the atomic structure, which is well-established on standard 
geometry relaxation methods, is, however, 
much more demanding from a first-principles
computational point of view and it is not
subject to scrutinity on a regular basis in molecular electronics.
Besides, there are  inherent uncertanties associated with the various
experimental procedures that can only be addressed by statistical analysis.

\begin{figure}
\includegraphics[width=2.6in]{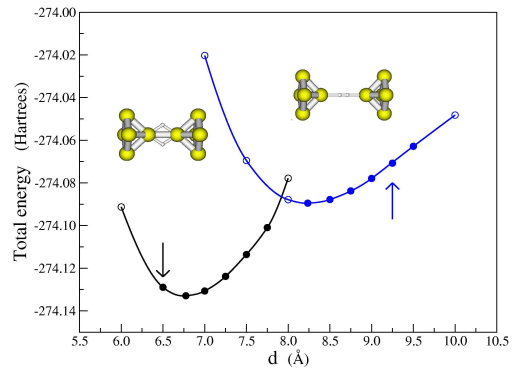}
\caption{Total energy of a 
H$_2$ molecule in a Pt nanocontact versus distance between
electrodes (outer planes). Blue circles correspond to the H atoms
positioned along the stretching axis ($z$) in molecular form 
(see right inset) and black circles
correspond to the H atoms positioned on the plane
perpendicular to the stretching axis (see left inset). Empty symbols
have been added as a guide to the eye which correspond to restricted 
relaxations where the H atoms are forced to stay on the $z$-axis (blue)
or on the plane perpendicular to it (black). Arrows indicate configurations 
for which vibrational mode frequencies closest to the experiment are obtained.
\label{PtH2}}
\end{figure}

In this paper we show how a combination of state-of-the-art
current-carrying first-principles electronic structure calculations
and a careful structural analysis can shed light on a recent 
experiment\cite{Smit:nature:02,Smit:thesis:03}
of electronic transport in Pt nanocontacts where, apparently,
H$_2$ molecules anchor themselves to 
the metal and modify the standard conductance histogram of Pt. 
A H$_2$ molecule is the simplest molecule 
and is ideal to carry out a detailed structural,
vibrational, and transport analysis. On the
basis of the extensive {\em ab initio} calculations we have performed, 
our main conclusion is the following: 
Contrary to the authors' explanation for the conductance
histograms where the main peak around $\mathcal G\approx \mathcal G_0$ 
(for $\mathcal G_0=2 e^2/h$) is attributed to a 
configuration where a H$_2$ molecule bridges the gap between Pt
electrodes (see also Ref.\ \onlinecite{Heurich:condmat:02}), 
we attribute this peak to a situation where the
H$_2$ molecule {\em dissociates at the nanocontact} and forms a stable
complex Pt$_2$H$_2$ where the two H atoms chemisorb
between electrode tip atoms [see left inset in Fig.\ \ref{PtH2}].
Our proposed structure  presents  vibrational modes in accordance with 
the reported inelastic effects\cite{Smit:nature:02} and with a more recent 
vibrational analysis\cite{Smit:thesis:03}.  
Finally, we speculate that the bridge with H in
molecular form could account for the peak around $\mathcal G  \approx
0.1 \mathcal G_0$ reported in Ref.\ 
\onlinecite{Smit:nature:02}, but not commented on there. 

\begin{figure}
\includegraphics[width=2.6in]{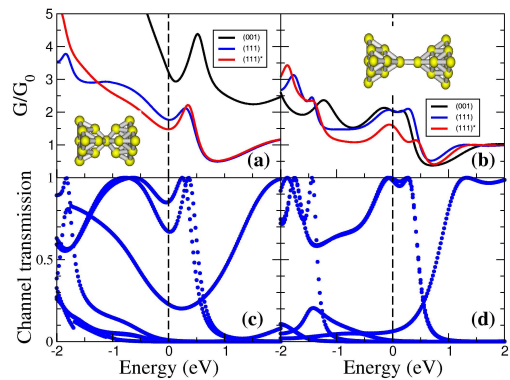}
\caption{(a) Conductance as a function of energy (referred to the Fermi
energy)  for the
(111)-oriented 6-3-1-3-6 clusters with (blue line) and without (red line)
mirror symmetry  and for the (100)-oriented  9-4-1-4-9 cluster; (b) 
same as in (a) but for the
(111)-oriented 6-3-1-1-3-6 clusters with (blue line) and without (red line)
mirror symmetry  and for the (100)-oriented  9-4-1-1-4-9 cluster; 
(c) channel transmission for the 6-3-1-3-6 cluster with mirror symmetry [see 
inset in (a)]; 
(d) same as in (c) but for the 6-3-1-1-3-6 cluster [see inset in (b)] 
with mirror symmetry.
\label{Pt-G}}
\end{figure}

\begin{figure}[b]
\includegraphics[width=2.6in]{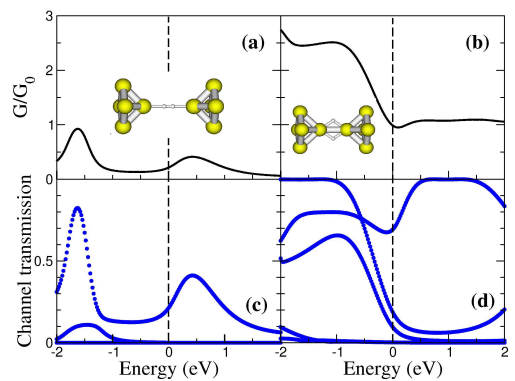}
\caption{(a) Conductance as a function of energy (referred to the Fermi
energy) for the  cluster  shown
in the inset where a H$_2$ molecule sits along the
stretching axis of the Pt nanocontact ($d=9.25$\AA). 
Dotted lines in (c) represent the transmission
of the individual eigenchannels for the cluster in (a). 
(b) Same as in (a) but for the
structure shown in the inset where the two H atoms form a double atomic
bridge ($d=6.5$\AA). 
Dotted lines in (d) represent the transmission of the individual eigenchannels
for the cluster in (b). The distance $d$ was chosen to give, in both cases,
the main vibrational mode frequency observed in the experiments. 
\label{PtH2-G}}
\end{figure}

For the structural and vibrational 
analysis of the Pt nanocontact with H we have
used the GAUSSIAN98 code\cite{Gaussian:98} and for the transport calculations 
we have employed the Gaussian Embedded Cluster Method (GECM),
recently developed by the authors\cite{Palacios:prb:01,
Palacios:prb:02,Louis:condmat:02} which, in turn, makes use of GAUSSIAN98.
The GECM consists of a first-principles evaluation at
the density functional theory (DFT) level of the density matrix and
the Fock matrix $F$ (or selfconsistent hamiltonian)  of 
the region determining the transport (here the H atoms and the
Pt electrode atoms in proximity to them) in a current-carrying situation. 
As regards
the DFT calculations for the core clusters (see insets in Figs.\ \ref{Pt-G} 
and \ref{PtH2-G}) we used the B3LYP functional\cite{Gaussian:98} , the basis sets and 
core pseudotentials described in Ref.\ \onlinecite{Pacios:jcp:85,
Hurley:jcp:86,Ross:jcp:90} for Pt and the cc-pVTZ basis for H\cite{Gaussian:98} .
Here we are concerned with the linear response conductance which
is calculated in the standard manner\cite{Agrait:pr:03}:
\begin{equation}
{\mathcal G}=\frac{2e^2}{h}{\rm Tr}[t^{\dagger}t].
\label{g}
\end{equation}
In this expression,  Tr denotes the trace over all the orbitals 
in the cluster. The matrix $t$ is given by 
\begin{equation}
t= \Gamma_{\rm R}^{1/2}G^{(+)}\Gamma_{\rm L}^{1/2}=
\left[\Gamma_{\rm L}^{1/2}G^{(-)} \Gamma_{\rm R}^{1/2} \right]^{\dagger},
\end{equation} 
where the Green function matrix (in orthogonal basis) is 
\begin{equation}
G^{(\pm)}=\left[(E\pm i\delta)1 - F-\Sigma^{(\pm)} \right ]^{-1},
\label{green}
\end{equation} 
while $1$ is the unity matrix.
The matrices $\Gamma_{\rm R}$ and $\Gamma_{\rm L}$
are given by $i(\Sigma^{(-)}_{\rm R}-\Sigma^{(+)}_{\rm R})$ and
$i(\Sigma^{(-)}_{\rm L}-\Sigma^{(+)}_{\rm L})$, respectively, where 
the self-energy $\Sigma^{(\pm)}=\Sigma^{(\pm)}_{\rm R}+\Sigma^{(\pm)}_{\rm L}$
represents the contribution of the semiinfinite left and right leads not 
included in the self-consistent 
evaluation\cite{Palacios:prb:01,Palacios:prb:02}.  
In order to single out the contribution of individual channels
to the current we diagonalize the matrix $t^{\dagger}t$.

As a starting point we consider the conductance of a clean Pt nanocontact
without H$_2$. It would be desirable to simulate the whole 
breaking(formation) process
of the nanocontact on stretching(indentation). 
This is, unfortunately, computationally not feasible from
first principles for Pt. Out of the many possible 
atomic arrangements, 
we have chosen to study the conductance for the following nanocontact
atomic configurations: 9-4-1-4-9 (single-atom contact) and  9-4-1-1-4-9
(chain contact)
for the (001) direction and  6-3-1-3-6 (single-atom contact)
and 6-3-1-1-3-6 (chain contact) for the (111) direction (the numbers indicate
the number of atoms per plane). We have considered two possible
plane orderings for the (111) direction: 6B-3A-$\dots$-3A-6B and 
6C-3A-$\dots$-3C-6A (remember that the bulk one 
is $\dots$ A-B-C-A-B-C $\dots$). 
These nanocontacts correspond to cases with and without mirror symmetry 
with respect to 
the middle plane perpendicular to the stretching axis $z$
[the (001) nanocontact has also been chosen with mirror symmetry]. 
Specifically, we have taken bulk atomic distances for the atoms in the
outer planes, but we have performed full relaxation in all the other
coordinates, including the distance between the two outer planes.
The resulting structures are thus representative of the ones responsible 
for the last(first) conductance plateau on
stretching(indentation) cycles. 

The conductance reaches $\approx 2\mathcal G_0$ at
the Fermi energy (set to zero) for the two short-chain
cases with mirror symmetry and drops slightly for the other one 
[see Fig.\ \ref{Pt-G}(b)]. Two conduction channels mainly contribute to the 
conductance  in the three clusters considered with the 
two-atom chain [see Fig.\ \ref{Pt-G}(d)]. Given the polycrystalline
nature of the electrodes, one should not expect mirror symmetry to be the
dominant situation in the experiment so we can expect an average conductance
around $\mathcal G\approx 1.75 \mathcal G_0$.
This value nicely coincides with the position of the lowest
conductance histogram peak reported in Ref.\ \onlinecite{Smit:nature:02}.
The conductance of 
the single-atom contacts presents similar values for the (111) direction 
(although, as Fig.\ \ref{Pt-G}(c) shows,
three channels now contribute to transport),
but is appreciable higher in the (001) direction [see Fig.\ \ref{Pt-G}(a)]. 
Although we cannot afford a statistical analysis,
the fact that the  position of the lowest conductance histogram peak
shifts to higher values when single-atom contacts are favoured (on 
only considering indentation cycles or higher bias voltages as reported 
in Ref.\ \onlinecite{Nielsen:condmat:02}) is consistent with our results.
The conductance is also here
noticeably dependent on the detailed atomic arrangement of the electrodes
(note that the conductance increases appreciably on the scale of 1 eV below
and above the Fermi energy). This is  due to the
fact that the Fermi level lies on the edge of the $d$ band for bulk and
the contact atom has almost bulk coordination.  This is also fairly
consistent with the fact that Pt histogram peaks are broad
in comparison to, e.g., those of Au\cite{Agrait:pr:03}.

When the same conductance measurements are performed in a H$_2$
atmosphere the histograms change noticeably. Inelastic vibrational
spectroscopy along with theoretical
support  lead the authors in Ref. \onlinecite{Smit:nature:02}
to conclude that a H$_2$ molecule can position itself along the
stretching axis of the Pt nanocontact right before it breaks apart. The
authors claim that this configuration 
accounts for the main histogram peak for conductance where 
$\mathcal G\approx \mathcal G_0$. With this claim in mind
we consider nanocontacts like that shown in the inset of
Fig.\ \ref{Pt-G}(b), 
but with a H$_2$ molecule anchored in between electrode tip atoms. 
The results presented below are for the smallest
cluster in the (001) direction, 4-1-H$_2$-1-4, 
but we have performed similar calculations
for (111) nanocontacts for reassurement. 
We maintain bulk interatomic distances for the four 
atoms in the two outer planes whose variable respective distance $d$ mimics
the effect of the piezo voltage.
For every $d$ we let the positioning of the H atoms
and that of the two contacting Pt atoms relax, starting from a configuration
where the H$_2$ molecule sits along the stretching axis. 
Full blue circles in Fig.\ \ref{PtH2} represent
the total energy vs. $d$ for all the stable configurations we have 
found in which the molecule stays on the $z$-axis (see right inset 
in Fig.\ \ref{PtH2}). Beyond
$d\approx 9.5$\AA\ the molecule sticks to one of the electrodes or 
dissociates. Below
$d\approx 8.25$\AA\ the  molecule tends to loose its alignment with
the $z$-axis. (As a guide to the eye we also present with empty circles
the energy for configurations 
where the molecule is forced to stay along the $z$-axis). In all these
points the H$_2$ retains its molecular form.
At the middle point ($d\approx 9$ \AA) the H-H bond 
distance and the H-Pt distance 
turn out to be similar to those proposed in Ref. \onlinecite{Smit:nature:02}.
Figure \ref{PtH2-G}(a) shows $\mathcal G(E)$ for $d=9.25$\AA \ (we chose the
distance that more closely reproduced the vibrational mode frequency reported 
in Ref.\ \onlinecite{Smit:nature:02}, see Table I). 
The conductance at the Fermi energy turns out
to be remarkably smaller than $\mathcal G_0$. A similar
calculation for the whole range of
stable configurations shown in Fig.\ \ref{PtH2} gave
a conductance in the range 0.2-0.5$\mathcal G_0$.
To our ease this result agrees with the intuitive picture:
Closed-shell molecules cannot conduct if the molecular character is
maintained after contact with the electrodes and the charge transfer
is much smaller than one. Both conditions apply here. The
charge transfer from the H$_2$ molecule to the electrodes is typically
$\approx 0.15$ or smaller
and the bonding and antibonding molecular levels maintain their character
since they are clearly visible
in the density of states (not shown here) around -6 eV and 17 eV, 
respectively. A value of 
conductance close to $\mathcal G_0$ is still possible  through 
hybridization of the Pt states with those of the 
molecule\cite{Heurich:condmat:02}. In our calculations, however,
this does not occur at the Fermi energy, but at $E \approx -1.5$ eV. 
We believe that this H$_2$ bridge actually accounts for a secondary 
peak around $\mathcal G\approx 0.1\mathcal G_0$ not discussed, but
clearly present in the reported data\cite{Smit:nature:02}. 

So, what accounts for the main conductance peak at $\approx \mathcal G_0$?. 
There is little doubt that this peak is H-related, but here
we propose an alternative to the H$_2$ molecular  bridge discussed above.
We now perform  calculations similar to the 
previous ones, but initially placing the molecule 
axis perpendicular to the stretching $z$-axis.  We let 
the Pt tip atoms and the H atoms relax towards equilibrium
for several values of $d$ {\em without assuming the H$_2$
to be in molecular form}.  The total energy of the stable configurations
we have found are represented by full black circles
in Fig.\ \ref{PtH2}. (Again, for completeness,
we also present results when the H atoms
are not allowed to leave the plane perpendicular to the $z$-axis). 
The two H atoms are 
only in molecular state at the upper limit of this curve ($d\approx 8$ \AA). 
For smaller values of $d$ the H$_2$ molecule breaks apart and the H
atoms position themselves up to 2.0 \AA\  apart from each other
forming a double atomic H bridge as the one shown in the left inset of 
Fig.\ \ref{PtH2}.  The results of conductance for  $d= 6.5$\AA\ (a distance
for which vibrational modes are 
closer to the experimental ones, see below and Table I)
are depicted in Fig.\ \ref{PtH2-G}(b). The conductance at the Fermi
energy approaches closely $\mathcal{G}_0$. 
The eigenchannel transmission results are also consistent
with the noise analysis in Ref.\ \onlinecite{Smit:nature:02} 
since, the fact that fluctuations do not strictly vanish at $\mathcal{G}_0$ may indicate
that more than one channel is contributing to the conductance.
At this point the distance between tip Pt atoms
is approximately that of bulk Pt and 
that between tip Pt atoms in Fig.\ \ref{Pt-G}(b). 
When comparing with the results presented in 
Fig.\ \ref{Pt-G}(b) we see that the H seems to block out one of the channels.
It is important to finally stress  here that the exact value 
of the conductance is not very dependent on the basis set
and the functional. However,
the atomic configuration of the electrodes in the cluster and the model used to 
represent the bulk of these electrodes introduces 
a source of uncertainty 
in the results. We believe that this is the source of the discrepancy
between our results and those presented in Refs. \onlinecite{Smit:nature:02} and
\onlinecite{Heurich:condmat:02}. After considering various other possibilities
in terms of cluster sizes and orientations, we can safely conclude that 
the stark difference between the conductance of the
two basic bridge structures studied in this work is robust.

We have also performed a full analysis of the H-H vibrational modes in a
configuration in which the H$_2$ axis is perpendicular to the $z$-axis
and for the distance that gives a conductance close to one quantum 
($d=6.5$\AA\, see Figs.\  \ref{PtH2} and \ref{PtH2-G}). 
The results for the mode frequencies are 
reported in Table I. The first mode shown in the
table (H atoms moving in opposite directions along the stretching
axis) is the one that has a frequency closer to the experimental data
(a frequency of 516$\pm$21 cm$^{-1}$ was reported in \cite{Smit:nature:02}).
The isotopic effect shows up clearly with frequencies
439 and 386 cm$^{-1}$ for H-D and D-D, to be compared with those obtained 
by multiplying the H-H frequency by the mass ratii (445 and 392 cm$^{-1}$,
respectively). It is worth mentioning that, although for D-D
the isotopic effects shows up for all modes, this is no longer true
for H-D. This mode induces a large variation of the
conductance (see Table I), again, an effect that is not the same 
for all modes. Note also that the mode with the two H atoms
moving in the same direction along the stretching axis, which also induces a 
significant change in the conductance (see second row in Table I),
has a frequency close to that of a second mode 
observed in the experiments (around 1200 cm$^{-1}$, see Ref.\ 
\onlinecite{Smit:thesis:03}). The latter is not reproduced by  
configurations with a molecular H$_2$ bridge (see Table I).

\begin{table}
\caption{
Frequencies (in cm$^{-1}$) of the six H-H vibrational modes in the
configuration for $d=6.5$\AA\, 
where the H$_2$ molecule dissociates and places itself with its axis 
perpendicular to the stretching axis (see Fig.\ \protect\ref{PtH2}).
Arrows indicate atomic displacements in the plane defined by H
and tip Pt atoms while $\pm$ denotes displacements perpendicular
to this plane. 
In the last column the numerical results for the absolute 
value of the relative change
in conductance induced by a 0.1 \AA \ displacement of the H
atoms are reported. The H-H vibrational modes for the configuration
with the H$_2$ molecule placed along the stretching axis 
and $d=9.25$\AA\, are shown in parenthesis.}

\label{Table}
\begin{tabular}{|c|c|c|c|c|}
\colrule
 mode   &  H-H  &  H-D  &  D-D & $ |\Delta G| /G $   \\
\colrule
$\leftarrow$  & 552 (4028) & 439  & 386  & 1.6\% \\
$\rightarrow$  &    &      &      &       \\
\colrule
$\leftarrow$  & 1315 (552) & 1155  & 934  & 0.82\% \\
$\leftarrow$  &    &      &      &       \\
\colrule
$\uparrow$  & 1380 (366) & 1036 & 979  & 0.36\% \\
$\uparrow$  &    &      &      &       \\
\colrule
$\uparrow$  & 1596 (1707) & 1508  & 1131 & 5.16\% \\
$\downarrow$  &    &      &      &       \\
\colrule
 +  & 250 (366) & 207  & 191  & 0.13\% \\
 +  &    &      &      &       \\
\colrule
  +  & 341 (1707) & 307  & 241  & 0.012\% \\
  --  &   &      &      &       \\
\colrule
\end{tabular}
\end{table}

In conclusion we have analyzed the conductance and vibrational modes of a
Pt nanocontact in the presence and absence of H. 
Our results for clean Pt agree with 
most data reported in the literature. On the contrary, our results
on Pt with H question the current interpretation of the experiments reported
in Ref.\ \onlinecite{Smit:nature:02}.
While H$_2$ seems to block the current, a novel complex
Pt$_2$H$_2$ formed at the neck
could be responsible for the reported histogram peak near 
$\mathcal G_0$. The vibrational modes of this complex also agree
with the experimental ones.

We acknowledge support from the 
Spanish Ministry of Science and Technology
under Grants No. 1FD97-1358 and No. MAT2002-04429-C03 and from
the Universidad de Alicante.
We thank C. Untiedt for sharing with us his point of view on the 
experimental results.


\end{document}